\documentclass[%
 reprint,
 groupedaddress,
 amsmath,amssymb,
 aps,
 pra,
 floatfix,
 twocolumn
]{revtex4-2}

\newcommand{\bra}[1]{\langle #1\rvert}
\newcommand{\ket}[1]{\lvert #1\rangle}

\newcommand{\op}[2]{\ket{#1} \bra{#2}}

\newcommand{\abs}[1]{\left\lvert#1\right\rvert}
\newcommand{\twomatrix}[4]{\left(\begin{array}{cc} #1 & #2 \\ #3 & #4 \end{array}\right)}

\DeclareMathOperator\erfc{erfc}
\DeclareMathOperator{\Tr}{Tr}

\usepackage{braket}
\usepackage{graphicx}% Include figure files
\usepackage{dcolumn}% Align table columns on decimal point
\usepackage{bm}% bold math

\usepackage{xcolor}

\usepackage{amsmath}
\usepackage{upgreek}
\usepackage{placeins}

\begin{document}

\preprint{APS/123-QED}

\title{Entropy transfer from a quantum particle to a classical coherent light field}

\author{John P. Bartolotta, Simon B. J\"ager, Jarrod T. Reilly, Matthew A. Norcia, James K. Thompson, Graeme Smith, and Murray J. Holland}
\affiliation{JILA, NIST, and Department of Physics, University of Colorado, 440 UCB, Boulder, CO 80309, USA}

\date{\today}

\pacs{Valid PACS appear here}

\begin{abstract}
In the field of light-matter interactions, it is often assumed that a classical light field that interacts with a quantum particle remains almost unchanged and thus contains nearly no information about the manipulated particles. To investigate the validity of this assumption, we develop and theoretically analyze a simple Gedankenexperiment which involves the interaction of a coherent state with a quantum particle in an optical cavity. We quantify the resulting alteration of the light field by measuring the fidelity of its initial and equilibrium states. Using Bayesian inference, we demonstrate the information transfer through photon measurements. In addition, we employ the concepts of quantum entropy and mutual information to quantify the entropy transfer from the particle to the light field. In the weak coupling limit, we validate the usually assumed negligible alteration of the light field and entropy transfer. In the strong coupling limit, however, we observe that the information of the initial particle state can be fully encoded in the light field, even for large photon numbers.
Nevertheless, we show that spontaneous emission is a sufficient mechanism for removing the entropy initially stored in the particle.
Our analysis provides a deeper understanding of the entropy exchange between quantum matter and classical light. 
\end{abstract}

\maketitle

%%%%%%%%%%%%%%%%%%%%%%%%%%%%%%%%%%%%%%%%%%%%%%%%%%%%%%%%%%%%%%
\section{Introduction}
%%%%%%%%%%%%%%%%%%%%%%%%%%%%%%%%%%%%%%%%%%%%%%%%%%%%%%%%%%%

When studying a quantum system, it is generally desirable for it to be prepared in a low-entropy configuration.
Such systems can populate a relatively small number of quantum states, which makes their dynamics more controllable, predictable, and comprehensible~\cite{Bloch,Zhang}.
If the system of interest is a gas consisting of atoms or molecules, its entropy is typically reduced by utilizing laser cooling and optical pumping techniques.
In these processes, the particles absorb light from an applied laser field and incoherently scatter the light into free space~\cite{MetcalfText}.
While these methods can irreversibly remove entropy from the gas, the second law of thermodynamics requires that the total entropy of the universe does not decrease, and therefore that the entropy of the gas must have been absorbed by some other system.
In this context, the most common explanation is that the entropy of the gas is absorbed by the vacuum modes of the quantized electromagnetic field~\cite{weisskopf}.
This process is often cast in the framework of open quantum systems whereby the quantized electromagnetic field is treated as an external reservoir, allowing for an irreversible reduction of the gas's entropy through the process of spontaneous emission~\cite{Esposito,Ptaszynski,Braunstein}.

It is typically explained that the reservoir can absorb a substantial amount of entropy due to the large number of possible emission configurations~\cite{MetcalfText,ketterleNobel}. 
Moreover, it is often stated that the coherent light field is not perturbed, and hence does not absorb entropy, because the quantum counterpart to the laser field, the coherent state, is unaffected by the absorption of photons by the gas.
However, there are some studies that predict entropy removal from the gas via interaction with the laser field~\cite{Korsunsky,Metcalf_2008,Metcalf_2015,Lignier,Miao}.
In an effort to further understand the underlying physics of laser cooling and optical pumping, we propose a simple Gedanken experiment that probes the change of a light field that coherently interacts with a particle containing nonzero initial entropy. 
Our results conclude that the entropy initially possessed by the particle can be imprinted on the applied light field, but the same information is also encoded in the spontaneous photons emitted by the particle.

%%%%%%%%%%%%%%%%%%%%%%%%%%%%%%%%%%%%%%%%%%%%%%%%%%%%%%%%%%%%%%
\begin{figure}
	\includegraphics[width=\linewidth]{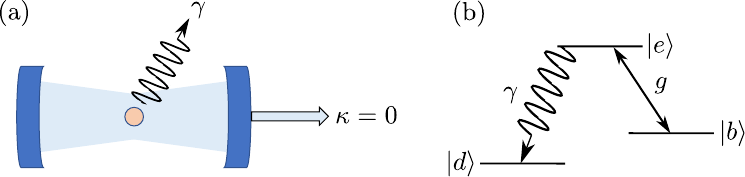}
	\caption{(a) A schematic of the experimental setup. A particle (circle) is placed in a lossless ($\kappa=0$) optical cavity that contains a coherent light field. The particle can undergo spontaneous emission into free space at rate $\gamma$. (b) Energy diagram of the particle's internal state structure. It is coupled to the cavity with strength $g$ on the bright transition ($\ket{e} \leftrightarrow \ket{b}$) and has a linewidth $\gamma$ on the dark transition ($\ket{e} \rightarrow \ket{d}$).}
	\label{fig:model}
\end{figure}
%%%%%%%%%%%%%%%%%%%%%%%%%%%%%%%%%%%%%%%%%%%%%%%%%%%%%%%%%%%%%%%

%%%%%%%%%%%%%%%%%%%%%%%%%%%%%%%%%%%%%%%%%%%%%%%%%%%%%%%%%%%
\section{Motivation}
%%%%%%%%%%%%%%%%%%%%%%%%%%%%%%%%%%%%%%%%%%%%%%%%%%%%%%%%%%%

We first consider the level of complexity necessary to demonstrate the state alteration and entropy transfer processes
with the goal of removing complications or details of any particular cooling or pumping scheme that might obscure the core physics of interest.
While other candidates may exist, we model the initial state of the laser field as a coherent state, as this is the quantum state that shares the most similarities with a classical coherent laser field~\cite{Agarwal}.
The coherent state is prepared in a lossless optical cavity, after which any external driving fields are turned off.
The cavity is an efficient way to incorporate quantization, and we anticipate that many of our findings also apply to free-space coherent light fields.
Because the essential physics of interest is the same, we focus only on the effects of optical pumping on the evolution of the particles' internal states.
To this end, we model the gas as a single, motionless particle that exists at an anti-node of the optical cavity.
With these considerations in mind, we construct the minimal experimental setup depicted in Fig~\ref{fig:model}(a).
The particle possesses two ground states and a single excited state, as shown in Fig~\ref{fig:model}(b).
The $\ket{b}\leftrightarrow\ket{e}$ (bright) transition is resonant with the optical cavity, which encapsulates the coherent interaction between the cavity field and the particle, while the $\ket{e}\rightarrow\ket{d}$ (dark) transition is mediated only by the spontaneous emission process, which models the incoherent interaction between the background radiation field and the particle.
The particle always transitions to the dark state $\ket{d}$ upon emission of a spontaneous photon so that the cavity is not trivially depleted of photons.

By studying this model, we aim to quantify the alteration of the cavity field due to its interaction with the particle and to determine if the particle's entropy is transferred to the cavity's degrees of freedom through their resulting correlations.
To achieve these goals, we employ both statistical and information theoretic techniques.
We first calculate the final cavity field state and determine its distinguishability from the initial cavity state qualitatively by comparing their Husimi $Q$-functions and quantitatively through their quantum fidelity.
Then, we demonstrate that information about the particle becomes encoded in the cavity field by using Bayesian inference.
Lastly, we quantify the amount of entropy transferred from the particle to the cavity field by calculating the quantum mutual information shared between the initial particle state and final cavity state.

%%%%%%%%%%%%%%%%%%%%%%%%%%%%%%%%%%%%%%%%%%%%%%%%%%%%%%%%%%%%%%
\section{Model}
%%%%%%%%%%%%%%%%%%%%%%%%%%%%%%%%%%%%%%%%%%%%%%%%%%%%%%%%%%%

The system of interest [Fig.~\ref{fig:model}(a)] consists of a three-level particle [Fig.~\ref{fig:model}(b)] coupled to a light field in a lossless ($\kappa=0$) optical single-mode cavity.
Let us denote the Hilbert spaces of the particle and light field as $A$ and $L$, respectively. 
In the time-independent interaction picture, the system is evolved according to the quantum master equation
\begin{equation}
\label{ME}
    \frac{d \hat \rho_{AL}}{dt} = 
    \frac{1}{i \hbar} \left[ \hat H_{AL} , \hat \rho_{AL} \right]
    + \gamma \mathcal{L}(\hat J) \hat \rho_{AL}.
\end{equation}
Here, the coherent particle-cavity interaction is described by the Jaynes-Cummings Hamiltonian
\begin{equation}
\label{ham}
    \hat H_{AL} = \frac{\hbar g}{2} \Big(\ket{b}\bra{e} \hat a^\dag +  \ket{e}\bra{b} \hat a\Big),
\end{equation}
where $g$ is the coupling strength, and $\hat{a}$ is the annihilation operator for the cavity field.
As is achieved in the moving frame of a particle in Doppler cooling, we have set the cavity to be resonant with the particle's bright state transition.
The environment surrounding the system, which we model as an infinite bandwidth bosonic bath that is coupled to the particle's dark transition, has been traced out under the Born-Markov approximation~\cite{Lindblad,Meystre}.
Its effects are incorporated through the Lindblad superoperator
\begin{equation}
    \mathcal{L}(\hat J) \hat \rho = 
    \hat J \hat \rho \hat J^\dag 
    - \frac{1}{2} \left(
        \hat J^\dag \hat J \hat \rho + \hat \rho \hat J^\dag \hat J 
    \right)
\end{equation}
with jump operator $\hat J = \ket{d} \bra{e}$. This term describes the spontaneous emission of photons into free space by the particle that occurs at a rate $\gamma$. We consider both analytical and numerical solutions to Eq.~\eqref{ME} in the sections that follow. We use both MATLAB and the QuantumOptics package in the Julia programming language for numerical calculations~\cite{MATLAB:2020b,JuliaQuantumOptics}.

To model the entropy initially possessed by the particle, it is prepared in the mixed state
\begin{equation}
\label{initParticle}
    \hat \rho_A(0) = x \ket{b}\bra{b} + (1-x)\ket{d}\bra{d},
\end{equation}
where $0\leq x\leq1$ is the probability that the particle begins in the bright state $\ket{b}$.
As already mentioned, the cavity field is initialized in a coherent state 
\begin{equation}
\label{coherentState}
\ket{\alpha} = e^{-|\alpha|^2/2} \sum_{n=0}^\infty \frac{\alpha^n}{\sqrt{n!}} \ket{n}, 
\end{equation} 
where $\ket{n}$ is the $n$-photon Fock state, so that its initial density matrix reads
\begin{equation}
\label{initCavity}
    \hat \rho_L(0) = \ket{\alpha} \bra{\alpha}.
\end{equation}

%%%%%%%%%%%%%%%%%%%%%%%%%%%%%%%%%%%%%%%%%%%%%%%%%%%%%%%%%%%
\begin{figure*}
	\includegraphics[width=\linewidth]{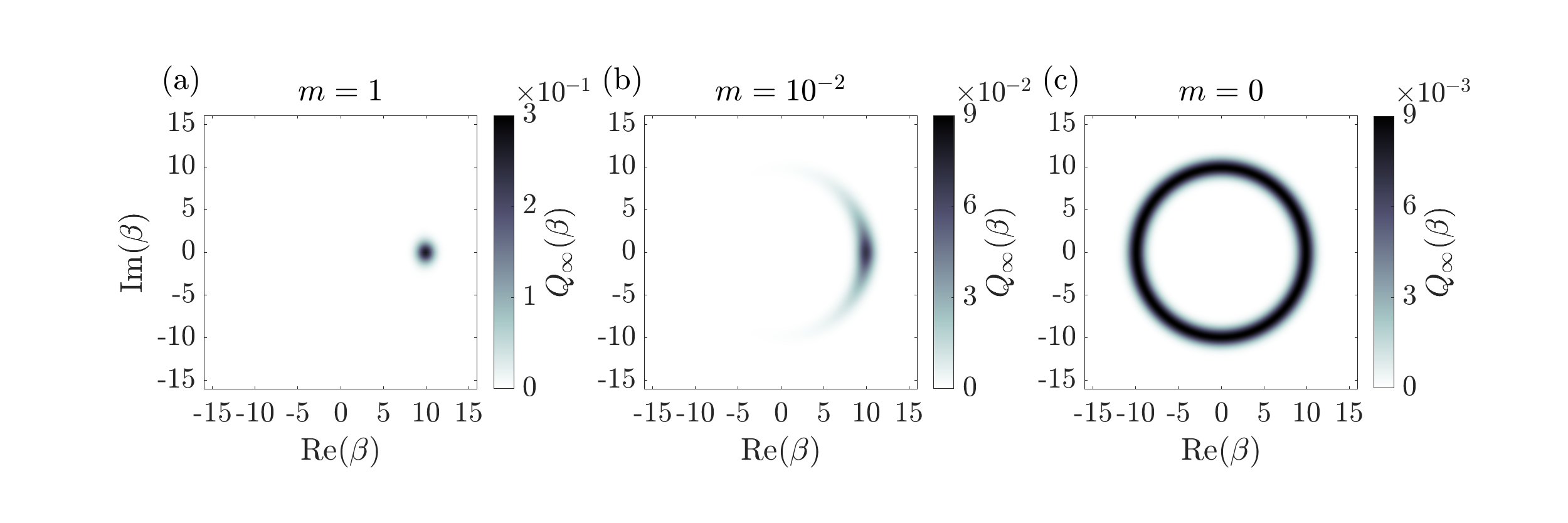}
	\caption{The $Q$-function of the final cavity state $Q_\infty(\beta)$ [see Eq.~\eqref{finalQ}] for various $m$. We have set $x=1$ and $\alpha = \sqrt{\bar n_0} = 10$. Although the amplitude of the state remains localized near $|\beta| = \sqrt{\bar n_0} = 10$, the spread in phase $\Delta \phi$ increases as $m$ decreases, indicating a change in the cavity state due to the particle-cavity interaction.}
	\label{fig:QFunction}
\end{figure*}
%%%%%%%%%%%%%%%%%%%%%%%%%%%%%%%%%%%%%%%%%%%%%%%%%%%%%%%%%%%

%%%%%%%%%%%%%%%%%%%%%%%%%%%%%%%%%%%%%%%%%%%%%%%%%%%%%%%%%%%
\section{Alteration of cavity state}
%%%%%%%%%%%%%%%%%%%%%%%%%%%%%%%%%%%%%%%%%%%%%%%%%%%%%%%%%%%

Here we demonstrate that the cavity state is altered due to its interaction with the particle.
But first, we address a common counterargument, which postulates that there should be no change to the cavity field because a coherent state is an eigenstate of the annihilation operator $\hat a$ by definition: $\hat a \ket{\alpha} = \alpha \ket{\alpha}$, and therefore is unperturbed by photon absorption followed by free-space spontaneous emission.
What this argument fails to consider is the stimulated emission of a photon back into the cavity, which can occur if the particle-cavity coupling rate is sufficiently large.
With this additional process in mind, we emphasize that a change of the cavity state is expected since it implies the action of the creation operator $\hat a^\dag$ [see Eq.~\eqref{ham}] on the coherent state, which yields nontrivial dynamics~\cite{AgarwalArticle}.

More specifically, we expect a change in the cavity state when the phase coherence between different Fock states in $\ket{\alpha}$ is scrambled.
This occurs when the relative phases of the relevant Fock state amplitudes become substantially altered, which we now characterize.
After an interaction time $t$, the accumulated phase for an $n$-photon Fock state $\ket{n}$ is $\phi_n = \sqrt{n} g t/2$.
If we approximate the interaction time by the excited state lifetime, $t \approx 1/\gamma$, then the relative accumulated phase $\Delta \phi$ between two Fock states $\ket{n}$ and $\ket{n+\Delta n}$ is
\begin{equation}
\label{relativePhase}
\Delta \phi (n, \Delta n )
= \phi_{n + \Delta n} - \phi_n
\approx \frac{g }{2 \gamma }
\left( 
  \sqrt{n + \Delta n}-\sqrt{n}
\right).
\end{equation}
For a coherent state $\ket{\alpha}$, which has initial average intracavity photon number $\bar n_0 = \langle \hat a^\dag \hat a(0) \rangle = |\alpha|^2$, the most relevant Fock states lie within the range $(\bar n_0 - \sqrt{\bar n_0},\bar n_0 + \sqrt{\bar n_0})$, in which $\sqrt{\bar n_0}$ is the variance of the photon distribution. Using Eq.~\eqref{relativePhase}, the relative accumulated phase between the central Fock state $\ket{\bar n_0}$ and the Fock states near to the edge of the coherent state $\ket {\bar n_0 \pm \sqrt{ \bar n_0}}$ is then
\begin{equation}
\label{relCoherent}
|\Delta \phi (\bar n_0, \sqrt{ \bar n_0} )|
\approx \frac{1}{4} \frac{g}{\gamma},
\end{equation}
in which we have assumed $\bar n_0 \gg 1$.
If the phase coherence is to be destroyed, then this relative phase must be much larger than unity. 
Defining the particle-cavity critical photon number $m \equiv \tfrac12 (\tfrac{\gamma}{g})^2$~\cite{Kimble} and using Eq.~\eqref{relCoherent}, phase scrambling of the coherent state is equivalent to the condition $m \ll 1$.
Therefore, $m$ is the important parameter for determining an alteration of the cavity state.

In Appendix~\ref{finalCavApp}, we derive an analytic expression for the final cavity state $\hat \rho_L(\infty)$ when the particle is initialized according to Eq.~\eqref{initParticle}.
It is parameterized by the initial bright state fraction $x$, initial intracavity photon number $\bar n_0$, and particle-cavity critical photon number $m$ [see Eq.~\eqref{finalCavity}].
To understand how the final cavity state differs from the initial coherent state, we pictorially compare their Husimi $Q$-functions, and then calculate their fidelity $F$. 

%%%%%%%%%%%%%%%%%%%%%%%%%%%%%%%%%%%%%%%%%%%%%%%%%%%%%%%%%%%%%%%%%%%%%%%%
\subsection{Cavity state $Q$-functions}
\label{QFunctionSection}
%%%%%%%%%%%%%%%%%%%%%%%%%%%%%%%%%%%%%%%%%%%%%%%%%%%%%%%%%%%%%%%%%%%%%%%%

To gain intuition for the differences between the initial and final cavity states, we calculate their Husimi $Q$-functions, which are defined as
\begin{equation}
\label{Qfunc}
    Q(\beta) \equiv \frac{\braket{\beta|\hat \rho |\beta}}{\pi}.
\end{equation}
Here, the coherent states $\ket{\beta}$ form a basis for the 2-dimensional optical phase space $(\text{Re}[\beta],\text{Im}[\beta])$.
The initial cavity state has $Q$-function
\begin{equation}
\label{initialQ}
    Q_0(\beta) = \frac{\braket{\beta|\hat \rho_L(0) |\beta}}{\pi}
    =\frac{|\braket{\alpha|\beta}|^2}{\pi}
    = \frac{1}{\pi} e^{-|\alpha - \beta|^2},
\end{equation}
while the $Q$-function for the final cavity state is
\begin{equation}
\label{finalQ}
    Q_\infty(\beta) = \frac{\braket{\beta|\hat \rho_L(\infty) |\beta}}{\pi}.
\end{equation}
The latter has a more complicated form since it is generally no longer a coherent state.
Figure~\ref{fig:QFunction} presents numerical plots of $Q_\infty(\beta)$ for various $m$.
To focus on the effects of the interaction, we choose $x=1$.
We also choose $\alpha = \sqrt{\bar n_0} = 10$ so that $Q_0(\beta)$ is localized and centered on the coordinate $(\text{Re}[\beta],\text{Im}[\beta])=(10,0)$ with a spread on the order of unity.

Figure~\ref{fig:QFunction}(a) displays $Q_\infty(\beta)$ for the intermediate-coupling case $m=1$.
We find that $Q_\infty(\beta)$ does not differ significantly from $Q_0(\beta)$, which indicates that the cavity field nearly remains in a coherent state.
In the case of stronger coupling [$m= 10^{-2}$, Fig.~\ref{fig:QFunction}(b)], however, $Q_\infty(\beta)$ remains localized near the circle $|\beta|=|\alpha|$, but is spread over a larger phase range $\Delta \phi$.
In the infinite-coupling limit $m\rightarrow0$ [Fig.~\ref{fig:QFunction}(c)], we find that $Q_\infty(\beta)$ has a uniform phase distribution, and therefore differs substantially from $Q_0(\beta)$.

Let us now interpret these results. 
The function $Q_\infty(\beta)$ always remains localized near the circle $|\beta|=|\alpha| = 10$ because the average number of photons in the cavity $\bar n = \braket{\hat a^\dag \hat a}$ does not significantly change.
This is a consequence of the particle's internal state structure [see Fig.~\ref{fig:model}(b)], which prevents a reduction of $\bar n$ by more than one, and our choice of a large initial intracavity photon number ($\bar n_0 = 100$).
On the other hand, we observe diffusion-like behavior of the phase $\phi$ as $m$ decreases because the particle undergoes more Rabi oscillations with each Fock state before emitting a spontaneous photon, thereby scrambling the cavity field's phase coherence.
The coherences vanish completely in the limit $m \rightarrow 0$, resulting in the uniform phase distribution shown in Fig.~\ref{fig:QFunction}(c).
This potentially extreme change in the phase distribution is largely responsible for the distinguishability of the initial and final cavity states, which we now quantify in terms of their fidelity.

%%%%%%%%%%%%%%%%%%%%%%%%%%%%%%%%%%%%%%%%%%%%%%%%%%%%%%%%%%%%%%
\subsection{Fidelity of initial and final cavity states}
\label{fidelitySection}
%%%%%%%%%%%%%%%%%%%%%%%%%%%%%%%%%%%%%%%%%%%%%%%%%%%%%%%%%%%%%%

We now quantify the alteration of the cavity field by calculating the Uhlmann-Jozsa fidelity
\begin{equation}
\label{uhlmannjozsa}
    F(\hat \rho, \hat \sigma)=
    \left(\Tr \sqrt{\sqrt{\hat \rho} \hat \sigma \sqrt{\hat \rho}}\right)^2
\end{equation}
between its initial and final states, which is a generalization of the transition probability for pure states~\cite{uhlmann,jozsa}. 
The fidelity satisfies $0 \leq F(\hat \rho, \hat \sigma) \leq 1$ for any two density matrices $\hat \rho$ and $\hat \sigma$, with $ F(\hat \rho, \hat \sigma)=1$ if and only if $\hat \rho = \hat \sigma$, and $F(\hat \rho, \hat \sigma)=0$ if $\hat \rho$ and $\hat \sigma$ have support on orthogonal subspaces.
In particular, if we find that $ F(\hat \rho, \hat \sigma) \neq 1$, then the cavity field is no longer in the coherent state $\ket{\alpha}$.

Because the cavity field is initially in a pure state, its fidelity with the final cavity state $\hat \rho_L(\infty)$ is simply
\begin{equation}
\label{simplifyFidelity}
    F \equiv F[\hat \rho_L(0), \hat \rho_L(\infty)] = \braket{\alpha | \hat \rho_L(\infty) | \alpha}.
\end{equation}
We point out that $F = \pi Q_\infty(\alpha)$, i.e., the fidelity is (up to a factor of $\pi$) the $Q$-function of the final cavity state evaluated at $\beta = \alpha$.
We can therefore qualitatively predict features of the fidelity $F$ through the plots in Fig.~\ref{fig:QFunction}.
For example, we expect $F$ to decrease as $m$ decreases due to the associated diffusion-like behavior of the $Q$-function.

\begin{figure}
	\includegraphics[width=\linewidth]{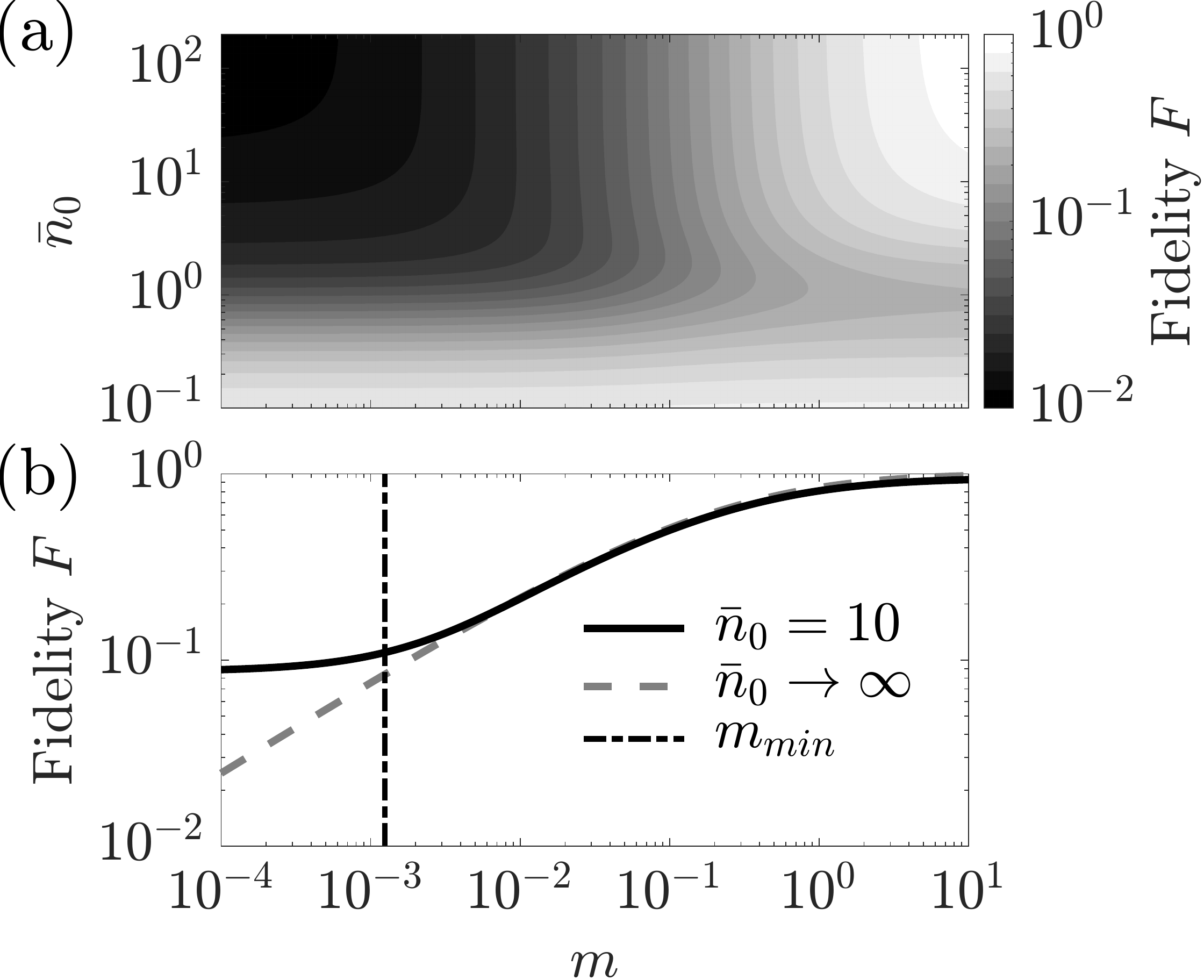}
	\caption{Fidelity $F$ between the initial and final cavity states $\hat \rho_L(0)$ and $\hat \rho_L(\infty)$ [Eq.~\eqref{uhlmannjozsa}] when the particle begins in the bright state [$x=1$, see Eq.~\eqref{fidGeneral}] as a function of the initial intracavity photon number $\bar n_0$ and the critical photon number $m$. (a) Contour plot of $F$, calculated by numerically evaluating Eq.~\eqref{simplifyFidelity}. When $m < 1$ and $\bar n_0 > 1$, we find that $F \ll 1$, signaling a significant change in the cavity state. (b) Numerical (solid, $\bar n_0 = 10$) and analytical [dashed, Eq.~\eqref{fidelityMDependence}] results for $F$ as a function of $m$. For $m<m_{min}$ [dot-dashed,  Eq.~\eqref{smallM}], the results diverge.}
	\label{fig:fidelity}
\end{figure}

The fidelity $F$ between the initial and final cavity state is calculated to be [see Eq.~\eqref{fidEarly}]
\begin{equation}
\label{fidGeneral}
    F =
    1-x\left[
        1 - f(\bar n_0,m)
    \right],
\end{equation}
in which $0<f(\bar n_0,m)<1$ is their fidelity conditioned on the particle being prepared in the bright state ($x=1$).
We present this conditional fidelity $f(\bar n_0,m)$ in Fig.~\ref{fig:fidelity}(a).
We find that $f \ll 1$, and therefore that the cavity state is significantly altered, when $\bar n_0 \geq 1$ and $m < 1$, which agrees with our qualitative $Q$-function analysis.
This region of parameter space corresponds to a cavity containing at least one photon (on average) and a strong particle-cavity coupling, respectively.

%%%%%%%%%%%%%%%%%%%%%%%%%%%%%%%%%%%%%%%%%%%%%%%%%%%%%%%%%%%%%%
%%%%%%%%%%%%%%%%%%%%%%%%%%%%%%%%%%%%%%%%%%%%%%%%%%%%%%%%%%%%%%

Because a typical laser field contains many photons, we focus on the thermodynamic limit $\bar n_0 \rightarrow \infty$ for the remainder of our investigation. 
In this limit, the conditional fidelity is calculated to be
\begin{equation}
\label{fidelityMDependence}
    \lim_{\bar n_0 \rightarrow \infty} f(\bar n_0,m)
     = \sqrt{2\pi m} \, e^{2m} \erfc(\sqrt{2m}),
\end{equation}
as shown in Appendix~\ref{fidelitySectionSM}. 
Figure~\ref{fig:fidelity}(b) displays Eq.~\eqref{fidelityMDependence} and a numerical result for $\bar n_0=10$, which agree very well when $m$ is sufficiently large:
\begin{equation}
\label{smallM}
    m > m_{min} \equiv \frac{1}{8 \pi^2 \bar n_0}.
\end{equation}
We now consider the strong ($m \ll 1$) and weak ($m \gg 1$) coupling limits.
From Eqs.~\eqref{fidGeneral} and~\eqref{fidelityMDependence},
\begin{equation}
\label{fidelityLimiting}
F\approx
\begin{cases}
    1-x(1-\sqrt{2 \pi m}), & m \ll 1\\
    \displaystyle 1 - \frac{x}{4m}, & m \gg 1
\end{cases}.
\end{equation}
Therefore, in the infinite-coupling limit $m \rightarrow 0$, the fidelity becomes $F=1-x$.
This shows that the cavity field can be altered by the interaction provided that the particle has a chance of starting in the bright state ($0 \ll x \leq 1$). 
However, in the zero-coupling limit $m\rightarrow\infty$, we find that $F=1$, and the field is not altered.

It is also interesting to consider how small the critical photon number must be for the initial and final cavity states to be substantially distinguishable, e.g., for their fidelity to be $F=\tfrac12$.
For simplicity and to achieve the greatest effect, we define such a critical photon number $m_{1/2}$ in the case when the particle is prepared in the bright state. 
From Eq.~\eqref{fidelityMDependence}, we find that this occurs when
\begin{equation}
\label{m12}
    m = m_{1/2} \approx 0.09.
\end{equation}
One can therefore observe a substantial alteration of the cavity state in an experimental setting by studying a system satisfying $m \leq m_{1/2}$, or more generally, $g \geq 2.3 \gamma \gg \kappa$.

These results are direct evidence that the cavity state can be altered by interaction with a particle, even in the limit of infinitely many photons, if the particle-cavity coupling is sufficiently large. We interpret the deviation from $F=1$ as result of the development of correlations between the particle and cavity states through the coherent interaction $\hat H_{AL}$ [see Eq.~\eqref{ham}].

%%%%%%%%%%%%%%%%%%%%%%%%%%%%%%%%%%%%%%%%%%%%%%%%%%%%%%%%%%%%%%
\section{Bayesian analysis}
\label{bayesSection}
%%%%%%%%%%%%%%%%%%%%%%%%%%%%%%%%%%%%%%%%%%%%%%%%%%%%%%%%%%%%%%

Now that we have demonstrated an alteration of the cavity state after interaction with the particle, we consider if any information about the particle is imprinted on the cavity field. 
As shown in Section~\ref{QFunctionSection}, the change in the cavity state manifests primarily in its phase.
This suggests that simply measuring the field intensity
\begin{equation}
    \langle \hat a^\dag \hat a(\infty)\rangle =
    \bar n_0 - x(1-e^{-\bar n_0})
\end{equation}
would not provide an effective way to extract this information, as it does not access the phase of the state.
To create a measurement that can distinguish differences in phase, we propose the following scheme. First, we displace the final cavity state according to the operation
\begin{equation}
\label{displacedCavityState}
    \hat \eta_L \equiv
    \hat D(-\alpha) \hat \rho_L(\infty) \hat D^\dag (-\alpha),
\end{equation}
which can be done, e.g., by removing a cavity mirror at $t\rightarrow\infty$ and feeding the final cavity state and a coherent state $\ket{\alpha}$ into separate ports of a beam splitter~\cite{paris}.
(We emphasize that any information about the particle contained in the cavity state is unaltered because this operation is unitary.)
After the displacement operation, we then perform photon number measurements on the resulting state $\hat \eta_L$.
To understand why this scheme provides us with cavity phase information, consider again the Husimi $Q$-function perspective of Section~\ref{QFunctionSection}.
Notice that the displacement operation [Eq.~\eqref{displacedCavityState}] simply shifts each phase space coordinate $\beta$ to the new coordinate $\beta - \alpha$.
Therefore, a nearly unperturbed cavity state [as in Fig.~\ref{fig:QFunction}(a)] would be shifted near to the phase space origin, so photon number measurements of $\hat \eta_L$ would yield low values, regardless of the high-noise photon statistics of the initial distribution.
Contrarily, only one point on the circle defining a significantly perturbed cavity state [as in Figure~\ref{fig:QFunction}(c)] would be mapped near to the phase space origin, so photon number measurements of $\hat \eta_L$ would typically yield much higher values.

To demonstrate the utility of our proposed measurement scheme, consider a situation wherein the particle is prepared in some diagonal mixed state [Eq.~\eqref{initParticle}], and we are tasked with determining the initial particle state (``start in $\ket{b}$" or ``start in $\ket{d}$") for a given experimental run by performing measurements exclusively on the displaced cavity state $\hat \eta_L$.
For simplicity, we calculate the results for the maximally mixed state ($x=\tfrac{1}{2}$), but our results can be generalized to any $x$.
By symmetry, the most successful approach without performing any measurements would be to sample from a flat probability distribution (the ``prior"): $P(\text{start in $\ket{b}$})=P(\text{start in $\ket{d}$})=\tfrac{1}{2}$, for which the probability that we would be correct is $P_\text{prior}(\text{correct})=\tfrac{1}{2}$.
What we show here is that a more accurate probability distribution can be constructed by incorporating the results from a single measurement of the displaced cavity photon number distribution $\braket{n | \hat \eta_L | n}$, thereby proving that information about the particle is present in the cavity field. 

For our purposes, it is sufficient to reduce the outcome space to a binary scenario: either we detect (i) zero photons ($n=0$, ``no click") or (ii) one or more photons ($n \geq 1$, ``click(s)").
This simplification is appropriate because we gain no additional information about the particle's initial state by distinguishing between nonzero numbers of clicks.
With this approach, we only need to calculate the vacuum state population $\braket{0|\hat \eta_L|0}$, which is equivalent to the fidelity $F$ [see Eq.~\eqref{fidGeneral}]. 
(Notice that one can experimentally probe $F$ by measuring this population.) 
The probability distribution in the event of ``no clicks" is then
\begin{align}
\label{likelihood}
    P(\text{no click}| \text{start in} \ket{i}) =
    \begin{cases}
    1, \;& i=d \\
    f(\bar n_0, m), \;&  i=b
    \end{cases},
\end{align}
where $f(\bar n_0,m)$ is the conditional fidelity. The ``click(s)" conditional probability distribution is complementary to Eq.~\eqref{likelihood}. 
With these results, we can use Bayesian inference to construct a posterior probability distribution for the initial particle state conditioned on the cavity measurement:
\begin{equation}
\label{bayes}
\begin{aligned}
    P(\text{start in} \ket{i}|C) \propto 
    P(&C|\text{start in} \ket{i}) P(\text{start in} \ket{i}); \\ 
    i \in \{b,d\}; \quad &C \in \{\text{click(s)}, \text{no click}\}.
\end{aligned}
\end{equation}
Here, $P(\text{start in} \ket{i})$ is the (initially flat) prior distribution for the initial particle state~\cite{holland}. 
The posterior probability distribution is
\begin{equation}
\label{posterior}
    \begin{aligned}
        P(\text{start in} \ket{d} |\, \text{click(s)}) & = 0; \\
        P(\text{start in} \ket{b} |\, \text{click(s)}) & = 1; \\
        P(\text{start in} \ket{d} |\, \text{no click}) & = \frac{1}{1+f(\bar n_0, m)}; \\
        P(\text{start in} \ket{b} |\, \text{no click}) & = \frac{f(\bar n_0, m)}{1+f(\bar n_0, m)}.
    \end{aligned}
\end{equation}

It is only left to demonstrate that the posterior probability distribution predicts the initial particle state more accurately than the prior probability distribution, i.e., that we would correctly predict the initial particle state with a probability satisfying $P_\text{post}(\text{correct})>P_\text{prior}(\text{correct})=\tfrac{1}{2}$, by sampling from Eqs.~\eqref{posterior}. 
Using the posterior probability distribution, the probability that we correctly predict the initial particle state is
\begin{equation}
\label{correct}
    P_\text{post}(\text{correct}) = \frac{1}{1 + f(\bar n_0, m)}.
\end{equation}
Since $0<f(\bar n_0, m)<1$, we find that $P_\text{post}(\text{correct}) \geq \frac{1}{2}$, and therefore conclude that information about the particle is present in the cavity field. 
We emphasize that we have increased our chance of predicting the initial particle state with a single measurement of the cavity state. 

We now focus on the thermodynamic limit $\bar n_0 \rightarrow \infty$, as often occurs in laser cooling and optical pumping. 
In this limit, the conditional fidelity $f$ is given by Eq.~\eqref{fidelityMDependence}. 
Using this form of $f$ in Eq.~\eqref{correct}, we find that 
\begin{equation}
\label{correctLimiting}
P_\text{post}(\text{correct}) \approx
\begin{cases}
    1- \sqrt{2 \pi m}, & m \ll 1\\
    \displaystyle \frac{1}{2}\left(1+ \frac{1}{8m}\right),  & m \gg 1
\end{cases}.
\end{equation}
In the infinite-coupling limit $m \rightarrow 0$, $P_\text{post}(\text{correct}) = 1$. Therefore, a single measurement of the cavity state can in principle predict the initial particle state with 100$\%$ accuracy, regardless of the high-noise photon statistics. 
This is possible because in this limit the final cavity states resulting from the particle starting in either $\ket{b}$ or $\ket{d}$ have orthogonal support, as evidenced by their vanishing fidelity \{see Eq.~\eqref{fidGeneral} and Ref.~\cite{Liang}\}.
In the zero-coupling limit $m \rightarrow \infty$, however, $P_\text{post}(\text{correct}) = \tfrac{1}{2} = P_\text{prior}(\text{correct})$, so the cavity measurement does not increase our chance of predicting the initial particle state.

%%%%%%%%%%%%%%%%%%%%%%%%%%%%%%%%%%%%%%%%%%%%%%%%%%%%%%%%%%%
\section{Mutual information}
%%%%%%%%%%%%%%%%%%%%%%%%%%%%%%%%%%%%%%%%%%%%%%%%%%%%%%%%%%%

In this section, we introduce an entropic perspective which we use to quantify the correlations between the particle, cavity field, and external reservoir due to the light-particle interactions.
In particular, we use this perspective to define the amount of entropy transferred from the particle to the cavity field.
Because the entropy of interest begins in the particle and ends in the cavity field, we posit that the amount of transferred entropy is characterized by the mutual information~\cite{Nielsen} shared between the initial particle state and the final cavity state.
In general, mutual information is defined as
\begin{equation}
\label{MIGeneral}
\begin{aligned}
I(Y:Z) & = S(Y) + S(Z) - S(YZ)\\
& = S(Y) - S(Y|Z),
\end{aligned}
\end{equation}
in which $Y$ and $Z$ are probability distributions, $\{S(Y),S(Z)\}$ are their information entropies, $S(YZ)$ is their joint entropy, and $S(Y|Z)$ is the conditional entropy of $Y$ given $Z$.
The entropy $S$ is given by the Shannon entropy 
\begin{equation}
\label{shannon}
    S(Y_\text{cl}) = - \sum_y P(y) \ln P(y)
\end{equation}
for a classical probability distribution $Y_\text{cl}$ with events $y$ and probabilities $P(y)$, whereas it is given by the von Neumann entropy
\begin{equation}
\label{vonNeumann}
    S(\hat \rho_Y) \equiv - \Tr \left[\hat \rho_Y\ln \hat \rho_Y\right]
\end{equation}
for a quantum probability distribution described by a density matrix $\hat \rho_Y$.
We have used the natural logarithm in Eq.~\eqref{shannon} for convenience.
Intuitively, $I(Y:Z)$ is equal to zero in the absence of any correlations between the two distributions, and is maximized if the higher-entropy distribution contains all of the information of the lower-entropy distribution:
\begin{equation}
\label{MIIneq}
\begin{aligned}
    0 \leq I(Y_\text{cl}:Z_\text{cl}) &\leq \text{min} [S(Y_\text{cl}),S(Z_\text{cl})], \\
    0 \leq I(\hat \rho_Y: \hat \rho_Z) &\leq 2 \, \text{min} [S(\hat \rho_Y),S(\hat \rho_Z)].
\end{aligned}
\end{equation}
Notice that the upper bound of quantum mutual information is twice as large as its classical counterpart~\cite{luo}.

To connect the formalism of this section with the Bayesian inference approach of Section~\ref{bayesSection}, we first calculate the amount of mutual information shared between the initial particle state and final cavity state as determined by the classical conditional probability distribution we derived through the photon number measurements [see Eqs.~\eqref{posterior}].
Then, we calculate their quantum mutual information from a density matrix approach, which incorporates all particle-cavity correlations. 
To understand the role of the cavity field in the entropy removal process, we also compare its final entropy to that of the reservoir, which contains the spontaneous photons emitted by the particle.

%%%%%%%%%%%%%%%%%%%%%%%%%%%%%%%%%%%%%%%%%%%%%%%%%%%%%%%%%%%
\subsection{Mutual information from cavity measurements}
%%%%%%%%%%%%%%%%%%%%%%%%%%%%%%%%%%%%%%%%%%%%%%%%%%%%%%%%%%%

Here we calculate the mutual information [Eq.~\eqref{MIGeneral}] between the initial particle state and final cavity state as determined by the photon number measurements of $\hat \eta_L$. 
Using the click probabilities from Eq.~\eqref{likelihood}, the notation of Eq.~\eqref{bayes}, and generalizing the conditional probability distribution in Eq.~\eqref{posterior} to any $x$, the conditional entropy of the initial particle state given the cavity measurements is
\begin{widetext}
\begin{equation}
\label{conditionalEntropy}
\begin{aligned}
    S[A(0)|L(\infty)] & =
    -\sum_C P(C) \sum_i P(\text{start in} \ket{i} | C) \ln [P(\text{start in} \ket{i} | C)] \\
    &\displaystyle = 
    -xf \ln \left[ \frac{fx}{1-x(1-f) }\right] 
     - (1-x) \, \ln \left[\frac{1-x}{1-x(1-f)}\right].
\end{aligned}
\end{equation}
\end{widetext}
From the classical mixture in Eq.~\eqref{initParticle}, the Shannon entropy of the initial state is
\begin{equation}
\label{particleEntropy}
 S[A(0)] \equiv S_0 = - x \ln x - (1-x) \ln (1-x).
\end{equation}
Together, Eqs.~\eqref{conditionalEntropy} and~\eqref{particleEntropy} can be used to calculate the (classical) particle-cavity mutual information
\begin{equation}
\label{MICavity}
    I[A(0):L(\infty)] = S[A(0)] - S[A(0)|L(\infty)].
\end{equation}

Because the cavity state is inherently quantum, the photon number measurements may not access all of its contained information.
Consequently, Eq.~\eqref{MICavity} underestimates the amount of mutual information shared between the particle and cavity.
However, as discussed in the next subsection, Eq.~\eqref{MICavity} will be useful for calculating the quantum mutual information in the thermodynamic limit $\bar n_0 \rightarrow \infty$, for which numerical calculations are intractable due to the infinite dimension of the cavity Hilbert space.

%%%%%%%%%%%%%%%%%%%%%%%%%%%%%%%%%%%%%%%%%%%%%%%%%%%%%%%%%%%
\subsection{Quantum mutual information}
%%%%%%%%%%%%%%%%%%%%%%%%%%%%%%%%%%%%%%%%%%%%%%%%%%%%%%%%%%%

We now include any additional correlations between the initial particle state $\hat \rho_A(0)$ and final cavity state $\hat \rho_L(\infty)$ by determining their quantum mutual information. 
As seen from Eq.~\eqref{MIGeneral}, this calculation requires knowing the joint distribution of these two states. 
Because these states are defined at different times, we calculate their joint distribution by incorporating an non-interacting auxiliary Hilbert space $R$ which purifies $\hat \rho_A(0)$ and hence contains the entropy of $\hat \rho_A(0)$ at all times.
The quantum mutual information will then be given by
\begin{equation}
\label{QMI}
    I(\hat \rho_R :\hat \rho_L) = S(\hat \rho_R) + S(\hat \rho_L) - S(\hat \rho_{RL}),
\end{equation}
in which all entropies $S$ are von Neumann entropies [see Eq.~\eqref{vonNeumann}].
In Appendix~\ref{entropyMISMSection}, we show that the density matrix $\hat \rho_{RL}$ is separable, which means that the quantum mutual information satisfies the classical inequality [Eq.~\eqref{MIIneq}(a)]
\begin{equation}
0 \leq I(\hat \rho_R :\hat \rho_L) \leq S_0,
\label{MIinequality}
\end{equation}
in which $S_0 \equiv S[\hat \rho_A(0)]$. In other words, somewhere between none [$I(\hat \rho_R :\hat \rho_L)=0$] or all [$I(\hat \rho_R :\hat \rho_L)=S_0$] of the entropy initially contained in the particle becomes encoded in the cavity field.
We elaborate on the repercussions of Eq.~\eqref{MIinequality} in Section~\ref{reservoirEntanglementSection}.

We now explain how to calculate $I(\hat \rho_R :\hat \rho_L)$. First, the initial particle state is purified through its Schmidt decomposition. 
That is, we view the particle ensemble as the reduced density matrix of a pure state,
$\hat{\rho}_A(0) = \Tr_R\op{u}{u}_{AR}$, where
\begin{equation}
\label{purification}
    \ket{u}_{AR} = \sqrt{x} \ket{b,b} + \sqrt{1-x} \ket{d,d}.
\end{equation}
It can be shown that the reduced density matrices $\hat \rho_A(0)$ and $\hat \rho_R$ have the same eigenvalues \cite{Nielsen}, which motivates the interpretation of $\hat{\rho}_R = \Tr_A\op{u}{u}_{AR}$ as an identically prepared ensemble to $\hat{\rho}_A (0)$, but with particles that are not interacting with the field. 
(This explains why the auxiliary particle describes the initial particle entropy: $S[\hat \rho_R(t)]=S_0$.)
The total system's density matrix becomes $\hat{\rho}_{ARL}$, and the master equation [Eq.~\eqref{ME}] is edited by incorporating the identity operator of $R$:
\begin{equation}
\label{substitutions}
\hat{H}_{AL} \rightarrow \hat{H}_{AL} \otimes \hat{\mathbb{I}}_R;  \qquad \hat J \rightarrow \hat J \otimes \hat{\mathbb{I}}_R.
\end{equation}
This updated master equation is then used to evolve the initial pure state
\begin{equation}
    \hat \rho_{ARL}(0) = \ket{u}\bra{u}_{AR} \otimes \ket{\alpha} \bra{\alpha}_L.
\end{equation}
We can then calculate the entropy of any subset of the $ARL$ composite Hilbert space by performing the appropriate trace operations.

\begin{figure}
\centerline{\includegraphics[width=\linewidth]{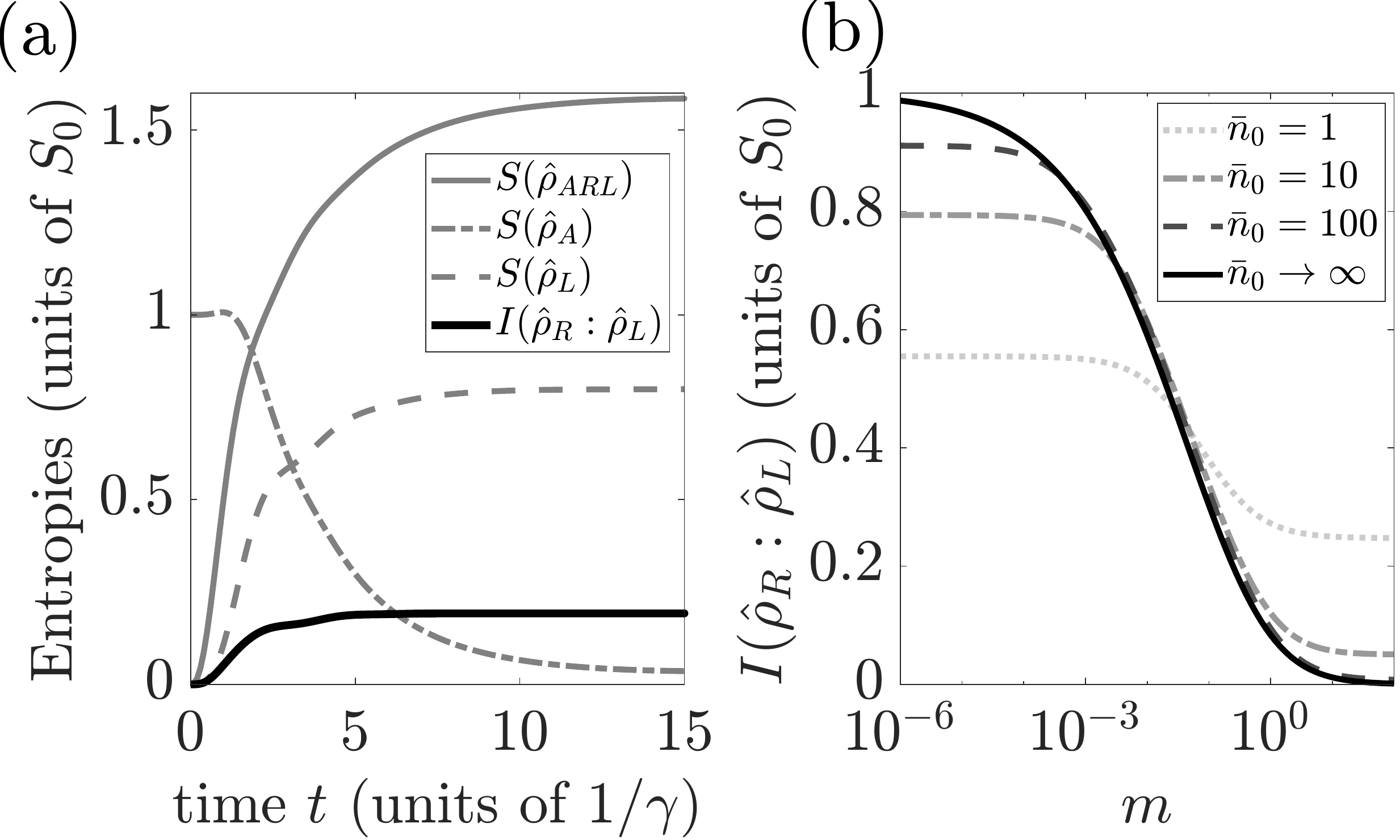}}
\caption{Entropic quantities scaled by the initial particle entropy $S_0 \equiv S[\hat \rho_A(0)]$. (a) von Neumann entropies $S[\hat \rho_M(t)]$ [see Eq.~\eqref{vonNeumann}] of various subspaces $M$ and quantum mutual information between the field and auxiliary particle $I(\hat \rho_R:\hat \rho_L)$ [see Eq.~\eqref{QMI}] as a function of time $t$. Parameters are: $x=m=0.5$ and $\bar n_0 = 5$. (b) Equilibrium $I(\hat \rho_R:\hat \rho_L)$ as a function of $m$ for various $\bar n_0$ with $x=0.5$. The $\bar n_0 \rightarrow \infty$ curve is given by Eq.~\eqref{MICavity}. The quantum mutual information reaches its maximum value $I(\hat \rho_R:\hat \rho_L)=S_0$ [see Eq.~\eqref{MIinequality}] in the thermodynamic ($\bar n_0 \rightarrow \infty$), infinite-coupling ($m \rightarrow 0)$ limit.}
\label{fig:I_S}
\end{figure}

Figure~\ref{fig:I_S}(a) presents various entropic quantities as a function of time $t$ for the choices $x=m=\tfrac{1}{2}$ and $\alpha = \sqrt{\bar n_0} = \sqrt{5}$. 
For convenience, we have scaled all quantities by the initial particle entropy $S_0$. 
As expected, the particle entropy $S[\hat \rho_A(t)]$ decreases (as occurs in laser cooling and optical pumping), the cavity state entropy $S[\hat \rho_L(t)]$ increases, and the entire $ARL$ space entropy increases as it evolves under the non-unitary dynamics. 
Importantly, the quantum mutual information $I(\hat \rho_ R:\hat \rho_L)$ increases and equilibrates to a nonzero value, which indicates the imprinting of the particle's initial entropy onto the cavity field.

We present numerical results for the equilibrium value of $I(\hat \rho_R:\hat \rho_L)$, scaled by $S_0$, as a function of $m$ for various $\bar n_0$ in Fig.~\ref{fig:I_S}(b) (non-solid curves).
We have dropped the time label $t \rightarrow \infty$ for notational simplicity.
We find for $\bar n_0 \geq 1$ that $I(\hat \rho_R:\hat \rho_L)$ can exceed $S_0/2$, and hence that a significant amount of information about the initial particle state can be imprinted on the cavity field, provided that $m$ is sufficiently small.
As $\bar n_0$ increases, we find numerically that the equilibrium quantum mutual information [Eq.~\eqref{QMI}] and mutual information as determined by the cavity measurements [Eq.~\eqref{MICavity}] converge. 
Therefore, we use Eq.~\eqref{MICavity} to analytically calculate the equilibrium value of $I(\hat \rho_R : \hat \rho_L)$ in the thermodynamic limit $\bar n_0 \rightarrow \infty$ (solid curve).
In the strong and weak-coupling limits, this simplifies to
\begin{flalign*}
    I(\hat \rho_R :\hat \rho_L) \approx&&
\end{flalign*}
\vspace{-2.1em}
\begin{equation}
\begin{cases}
    \displaystyle S_0 - \epsilon(x,m), & \displaystyle m \ll \text{min} \left[1, \, \frac{1}{2 \pi} \left( \frac{1-x}{x}\right)^2\right]\\
    \displaystyle -\frac{x\ln x}{4m}, & m \gg 1
\end{cases} \;,
\end{equation}
in which
\begin{equation}
    \epsilon(x,m) = - \sqrt{2 \pi m} 
    \left[
    \ln \sqrt{2 \pi m} + \ln \left( \frac{x}{1-x}  \right)
    -1
    \right]
\end{equation}
and $0 \leq \epsilon \ll S_0$.
The additional constraint on $m$ in the strong-coupling limit is a consequence of the highly nonlinear behavior of $I(\hat \rho_R:\hat \rho_L)$ for small $m$.
In the zero-coupling limit $m \rightarrow \infty$, we find that $I(\hat \rho_R:\hat \rho_L) = 0$, so the cavity field does not contain the information entropy of the initial particle state, as expected.
However, in the infinite-coupling limit $m \rightarrow 0$, we find that $I(\hat \rho_R:\hat \rho_L) = S_0$, which means that the cavity field contains complete information about the initial particle state.
This agrees with our Bayesian inference result in the previous section. 
We rigorously prove that $I(\hat \rho_R:\hat \rho_L)=S_0$ only when $\bar n_0 \rightarrow \infty$ and $m\rightarrow0$ in Appendix~\ref{entropyMISMSection}.

%%%%%%%%%%%%%%%%%%%%%%%%%%%%%%%%%%%%%%%%%%%%%%%%%%%%%%%%%%%%%%
\subsection{Entanglement with the reservoir}
\label{reservoirEntanglementSection}
%%%%%%%%%%%%%%%%%%%%%%%%%%%%%%%%%%%%%%%%%%%%%%%%%%%%%%%%%%%

In the previous sections, we demonstrated in several ways that the final cavity state contains information about the initial particle state.
It is therefore tempting to conclude that the cavity field has removed entropy from the particle.
However, we have yet to consider the entropy contained in the only remaining subspace: the external reservoir, which contains the spontaneous photons emitted by the particle.
If we denote the reservoir Hilbert space by $P$, it can be shown that [see Eq.~\eqref{spontEntropyBigger}]
\begin{equation}
\label{photonEntropy}
    S[\hat \rho_P(\infty)] \geq S[\hat \rho_L(\infty)].
\end{equation}
Physically, Eq.~\eqref{photonEntropy} demonstrates that the spontaneous photons always contain at least as much information about the initial particle state as the cavity field.
In this sense, spontaneous emission is a sufficient mechanism for removing entropy from the particle.
A specific instance of this inequality can be seen in Fig.~\ref{fig:I_S}(a) by noticing that $S[\hat \rho_{ARL}(t)]=S[\hat \rho_P(t)]$.

Although the initial particle state and final cavity state are correlated, the separability of $\hat \rho_{RL}$ indicates that they are not entangled.
This is why their quantum mutual information satisfies the stricter, classical bound [see Eqs.~\eqref{MIIneq} and~\eqref{MIinequality}].
However, Eq.~\eqref{photonEntropy} can be used to show that both the particle and the cavity field become entangled with the external reservoir. 
As shown in Appendix~\ref{photonEntropySM},
\begin{equation}
\label{reservoirEntanglement}
\begin{aligned}
    I[\hat \rho_P(\infty) : \hat \rho_R(\infty)] \geq S[\hat \rho_R(\infty)], \\
    I[\hat \rho_P(\infty) : \hat \rho_L(\infty)] \geq S[\hat \rho_L(\infty)].
\end{aligned}
\end{equation}
Equations~\eqref{reservoirEntanglement} imply entanglement when the inequalities are strict, which occurs if $m>0$. Consequently, the particle and cavity can develop richer quantum correlations with the reservoir than with each other.

%%%%%%%%%%%%%%%%%%%%%%%%%%%%%%%%%%%%%%%%%%%%%%%%%%%%%%%%%%%%%%
\section{Conclusion}
%%%%%%%%%%%%%%%%%%%%%%%%%%%%%%%%%%%%%%%%%%%%%%%%%%%%%%%%%%%

We have demonstrated through a simple Gedankenexperiment under what conditions the entropy of a quantum system can be imprinted on a classical coherent light field, which we modeled as a coherent state in a lossless cavity.
We have quantified the cavity state's alteration due to the particle-cavity interaction through the measurement of fidelity and shown that the cavity field contains information about the initial state of the particle by the method of Bayesian inference and using quantum information theoretic techniques. 
Our results demand reconsideration of the underlying physics of laser cooling and optical pumping in the strong particle-light coupling regime
~\cite{swap_exp,swap_theory,SWAP_MOT_theory,corder2015laser,Metcalf_2008}, as it suggests that the assumption of an unperturbed light field is not necessarily accurate.

The entropy transfer from the particle to the light field could be realized in an experimental setting by studying a system that satisfies $\kappa \ll \gamma, g$. Although we have mainly focused on the high-$\bar n_0$ limit, we believe that there is also interesting physics in the low-$\bar n_0$ regime. 
Our model could also be generalized to further understand some cavity-based quantum memories~\cite{impedanceMatched,spinWave,cavityEnhancedMemory,quantumMem1,quantumMem2,Giannelli_2018} by, e.g., incorporating cavity pumping and loss ($\kappa \neq0$) or initializing the cavity field in a different quantum state.
We also anticipate that this system can be used as a platform to generate and study novel photon-subtracted states~\cite{Agarwal}.

There are many ways to extend this study for the purpose of modeling a laser cooling process more accurately. 
Most notably, the incorporation of cavity pumping and loss could permit a nontrivial equilibrium solution even when the particle can relax back to the bright state, which would bring the particle's internal state structure closer to that of typical two-level models. 
In this case, one could altogether remove the effects of the background radiation field on the particle to further investigate the entropy dynamics of the laser-particle system~\cite{cavityCooling}. 
Of course, particle motion could also be incorporated, potentially allowing for another Hilbert space to exchange entropy.

One could also study this system in the context of phase space compression. 
However, the rich correlations generated by quantum mechanical processes preclude a clear phase space approach for quantum systems~\cite{Oliva,Bernardini,Bernardini2}. 
This would require, for example, a deeper understanding of the connection between Wigner trajectories and Liouville's theorem~\cite{Sala}.
On a related note, one could consider the use of quantum discord as a measurement of quantum correlation~\cite{Ferraro} as opposed to von Neumann entropy.

%%%%%%%%%%%%%%%%%%%%%%%%%%%%%%%%%%%%%%%%%%%%%%%%%%%%%%%%%%%
\section{Acknowledgments}
%%%%%%%%%%%%%%%%%%%%%%%%%%%%%%%%%%%%%%%%%%%%%%%%%%%%%%%%%%%

The authors thank John Cooper, Konrad Lehnert, and Vera Sch\"afer for helpful discussions and comments.
This work was supported by NIST and by NSF Grant No. PHY 1806827, NSF PFC Grant No. PHY 1734006, and by NSF Grant No. OMA 2016244.

\appendix
%%%%%%%%%%%%%%%%%%%%%%%%%%%%%%%%%%%%%%%%%%%%%%%%%%%%%%%%%%%%%%%
\section{Final cavity field state}
\label{finalCavApp}
%%%%%%%%%%%%%%%%%%%%%%%%%%%%%%%%%%%%%%%%%%%%%%%%%%%%%%%%%%%%%%%%

In order to derive an analytic form of the cavity field state in the limit $t\rightarrow \infty$, we separate the density matrix into bright and dark manifolds:
\begin{equation}
    \begin{aligned}
\hat{\rho}_b =& \braket{b |\hat{\rho} |b} \op{b}{b} + \braket{b|\hat{\rho}|e} \op{b}{e}\\
&+ \braket{e|\hat{\rho}|b} \op{e}{b} + \braket{e|\hat{\rho}|e} \op{e}{e},\\
\hat{\rho}_d =& \braket{d|\hat{\rho} |d} \op{d}{d}.
    \end{aligned}
\end{equation}
This simplifies the formalism because the initial particle-cavity state
\begin{equation}
\label{initialState}
\hat{\rho}(0) = \left[ x \op{b}{b} + (1-x) \op{d}{d} \right] \otimes \op{\alpha}{\alpha} \\
\end{equation}
explicitly assumes there are no coherences between the three internal states. Taking a time derivative and using the master equation [Eq.~\eqref{ME}] in the form
\begin{equation}
    \frac{d\hat{\rho}}{dt} = \frac{1}{i\hbar} \left( \hat{H}_{\text{eff}} \hat{\rho} - \hat{\rho} \hat{H}_{\text{eff}}^{\dagger} \right) + \hat{\mathcal{L}}_{\text{jump}}\left( \hat{\rho} \right),
\end{equation}
we find
\begin{equation}
\label{editedMaster}
    \dot{\hat{\rho}}_d = \gamma \braket{e|\hat{\rho}|e} \op{d}{d}, \quad \dot{\hat{\rho}}_b = \frac{1}{i\hbar} \left[ \hat{H}_{\text{eff}} \hat{\rho}_b - \hat{\rho}_b \hat{H}_{\text{eff}}^{\dagger} \right].
\end{equation}
Here, we used $\bra{i} \hat{\mathcal{L}}_{\text{jump}} \ket{j}=0$ for $i,j \in \{ b,e \}$ and defined
\begin{equation}
\begin{aligned}
    \hat{H}_{\text{eff}} &= \hat{H} - \frac{i \hbar \gamma}{2} \op{e}{e},\\ \hat{\mathcal{L}}_{\text{jump}}(\hat{\rho}) &= \gamma \braket{e|\hat \rho |e} \op{d}{d}.
\end{aligned}
\end{equation}

We first diagonalize the non-Hermitian matrix $\hat H_\text{eff}$ in the subspace $\{ \ket{e,n-1}, \ket{b,n} \}$ 
\begin{equation} \label{Hn}
    \hat{H}_n = \frac{\hbar}{2}\twomatrix{-i \gamma}{g \sqrt{n}}{g \sqrt{n}}{0}
\end{equation}
for $n \geq 1$.
The zero-excitation subspace $\{\ket{g,0}\}$ is constant in time.
The eigenvalues are $\lambda_{\pm}^{(n)} = -i\hbar \Lambda_{\pm}^{(n)}$, with 
\begin{equation}
    \Lambda_{\pm}^{(n)} \equiv \frac{\gamma}{4} \pm \frac{i}{2} \sqrt{ng^2 - \frac{\gamma^2}{4}},
\end{equation}
and the kernel of the matrix $\hat{H}_n - \lambda_{\pm}^{(n)} \hat{\mathbb{I}}_2$ is spanned by
\begin{equation}
    \hat{V}_n = \frac{\hbar}{c_n} \twomatrix{-i \Lambda_+^{(n)}}{-\frac{g \sqrt{n}}{2}}{\frac{ g \sqrt{n}}{2}}{-i \Lambda_+^{(n)}}
\end{equation}
with normalization factor $c_n^2 = \left(-i\hbar \Lambda_+^{(n)} \right)^2 + \hbar^2 g^2 n/4$.
Here, $\hat{\mathbb{I}}_2$ is the 2$\times$2 identity matrix.
We now define $\hat{V} = \sum_n \hat{e}_{n,n} \otimes \hat{V}_n$, in which $\hat e_{k,l}$ is a single-entry matrix with a 1 at position $(k,l)$ and zeroes elsewhere, so that the bright manifold matrix can be written as $\hat{\rho}_b = \hat{V} \hat{\rho}_V \hat{V}^{\dagger}$.

We can use the method of Laplace transforms to solve for the transformed density matrix:
\begin{equation} \label{LaplaceEq}
    s L[\hat{\rho}] = L [ \dot{\hat{\rho}}](s) + \hat{\rho}(0),
\end{equation}
which we then can solve for steady-state:
\begin{equation} \label{SteadyStateLaplace}
    \hat{\rho} (t \rightarrow \infty) = \lim_{s \rightarrow 0} L [ \dot{\hat{\rho}}](s) + \hat{\rho} (0).
\end{equation}
Projecting with $\bra{d}$ and $\ket{d}$ and using Eq.~\eqref{editedMaster}, $\braket{d | \hat{\rho}_b |d}=0$, and $\braket{e|\hat{\rho}_d | e}=0$, we find 
\begin{equation} \label{dSteadyStateLaplace}
    \hat{\rho}_d (t \rightarrow \infty) = \gamma \lim_{s \rightarrow 0} \braket{e|L [ \hat{\rho}_b](s)|e} + \braket{d |\hat{\rho}_d (0) |d}.
\end{equation}
To solve for the first term on the right-hand side of Eq.~\eqref{dSteadyStateLaplace}, we first find
\begin{equation}
\begin{aligned}
    L [\hat{\rho}_V] (s) &= L \left[ \hat{V}^{-1} \hat{\rho}_b (\hat{V}^\dagger)^{-1} \right] (s) \\
    &= \sum_{n,n' \geq 1} \hat{e}_{n',n} \otimes \hat{R}_{n',n}(s),
\end{aligned}
\end{equation}
with
\begin{equation}
\begin{aligned}
    \hat{R}_{n',n}&(s) = \frac{\hbar^2}{c_{n'} c_n^*} \braket{n'|\alpha} \braket{\alpha|n} \\
    &\times \twomatrix{\frac{ g^2 \sqrt{nn'}}{4 (s+ \Lambda_+^{(n')} + \Lambda_+^{(n)*})}}{\frac{i g \sqrt{n'}\Lambda_+^{(n)*}}{2 (s+ \Lambda_+^{(n')} + \Lambda_-^{(n)*})}}{-\frac{i g \sqrt{n}\Lambda_+^{(n')}}{2 (s+ \Lambda_-^{(n')} + \Lambda_+^{(n)*})}}{\frac{ \Lambda_+^{(n')} \Lambda_+^{(n)}}{(s+ \Lambda_-^{(n')} + \Lambda_-^{(n)*})}} .
\end{aligned}
\end{equation}
We next calculate 
\begin{equation}
\label{laplaceBright}
    L [ \hat{\rho}_b ](s) = \sum_{n,n' \geq 1} \hat{e}_{n',n} \otimes  \hat{V}_{n'} \, \hat{R}_{n',n}(s) \, \hat{V}_n^{\dagger}.
\end{equation}
Projecting Eq.~\eqref{laplaceBright} onto the excited state subspace and incorporating the initial condition from Eq.~\eqref{initialState}, Eq.~\eqref{dSteadyStateLaplace} becomes
\begin{equation}
\label{rhoDEq}
\begin{aligned}
\hat{\rho}_d (\infty) = &\op{d}{d} \otimes \biggl[ 
(1-x) \op{\alpha}{\alpha} \\
&+ x |\alpha^2| 
\sum_{l,l' \geq 0} K_{l,l'} \braket{l'|\alpha} \braket{\alpha|l} \op{l'}{l}
\biggr],
\end{aligned}
\end{equation}
with
\begin{equation}
\label{K}
    K_{l,l'} \equiv \left( 1 + \frac{l+l'}{2} + \frac{(l-l')^2}{8 m} \right)^{-1},
\end{equation}
in which $m \equiv \tfrac12 (\tfrac{\gamma}{g})^2$ is the particle-cavity critical photon number~\cite{Kimble}.

Projecting $\bra{b}$ and $\ket{b}$ on Eq.~\eqref{SteadyStateLaplace} and noting that the steady-state bright state population results from the cavity being in the vacuum state $\ket{0}$, we have
\begin{equation}
\label{rhoBEq}
    \hat{\rho}_b (\infty) = x e^{-\abs{\alpha}^2} \op{b}{b} \otimes \op{0}{0}.
\end{equation}
Tracing over the atomic states, we find that the final cavity field state is
\begin{equation}
\label{finalCavity}
    \hat{\rho}_L (\infty) = (1-x) \ket{\alpha}\bra{\alpha} + x \hat \rho_c,
\end{equation}
in which
\begin{equation}
\label{rhoc}
\begin{aligned}
\hat \rho_c = &e^{-\abs{\alpha}^2} \op{0}{0}\\
&+ \abs{\alpha}^2 \sum_{l,l' \geq 0} K_{l,l'}  \braket{l'|\alpha} \braket{\alpha|l} \op{l'}{l}
\end{aligned}
\end{equation}
characterizes the effects of the interaction.

%%%%%%%%%%%%%%%%%%%%%%%%%%%%%%%%%%%%%%%%%%%%%%%%%%%%%%%%%%%%%%%%%%%%%%%%
\section{Fidelity between initial and final cavity field states}
\label{fidelitySectionSM}
%%%%%%%%%%%%%%%%%%%%%%%%%%%%%%%%%%%%%%%%%%%%%%%%%%%%%%%%%%%%%%%%%%%%%%%%

Here we calculate the fidelity
\begin{equation}
\label{fidelitySM}
    F(\hat \rho, \hat \sigma) = \left( \Tr \sqrt{\sqrt{\hat \sigma} \hat \rho \sqrt{\hat \sigma}} \right)^2
\end{equation}
between the initial and final states of the cavity. As seen from Eq.~\eqref{finalCavity},
\begin{equation}
\label{fidEarly}
\begin{aligned}
     F \bigl[ \hat \rho_L(\infty),\hat \rho_L(0) \bigr] & =
     F \bigl[ \hat \rho_L(\infty),\ket{\alpha}\bra{\alpha}]\\
     &=\braket{\alpha | \hat \rho_L(\infty) | \alpha } \\
     &= 1-x\left[1 - f(\bar n_0,m)\right],
\end{aligned}
\end{equation}
in which
\begin{equation}
\label{conditionalFidelity}
    f(\bar n_0,m) \equiv e^{-2 \bar n_0} 
    \left(
    1 +
    \bar n_0
    \sum_{l,l'\geq 0} K_{l,l'} \frac{\bar n_0^{l+l'}}{l!l'!}
    \right)
\end{equation}
is the fidelity conditioned on the particle beginning in the bright state, and $\bar n_0 \equiv \braket{\hat a^\dagger \hat a(0)}$. We use this result to create a numerical contour plot of $F$ in Fig.~\ref{fig:fidelity}(a). 

We now calculate an analytic expression for $F$ in the thermodynamic limit $\bar n_0 \rightarrow \infty$. Using the Laplace transform of $K_{l,l'}$ [see Eq.~\eqref{K}] and the Fourier transform of the resulting Gaussian term to reduce its order, we find that the sum in Eq.~\eqref{fidEarly} can be rewritten as
\begin{equation}
\label{fidSum}
    \sum_{l,l'\geq0} K_{l,l'} 
        \frac{\bar n_0^{l+l'}}{l!l'!} = 
    \int_{-\infty}^\infty 
    \int_0^\infty ds \,
     du \,
    G(u,s),
\end{equation}
in which
\begin{equation}
    G(u,s) = \sqrt{\frac{2 m}{\pi s}}
    e^{-s} 
    e^{2 \bar n_0 e^{-s/2}  \cos u}
    e^{- 2 m u^2/s}.
\end{equation}
In the limit $\bar n_0 \gg 1$, $G(u,s)$ is localized around $s=0$ and $u=2 \pi p$, where $p$ is an integer. This can be seen from the dominating term $\exp \left(2 \bar n_0 e^{-s/2} \cos u\right)$. Expanding the argument of this term to lowest order in $s$ and $u$, we find that
\begin{equation}
\label{approxLargeNBar}
    e^{2 \bar n_0 e^{-s/2} \cos u} \approx
   e^{2 \bar n_0} e^{- s \bar n_0 }  \sum_{p \in \mathbb{Z}}  e^{- \bar n_0 (u- 2 \pi p)^2} .
\end{equation}
From this result, we see that the relevant domains are $0 < s < 1/\bar n_0$ and $|u - 2 \pi p|< 1 / \sqrt{\bar n_0}$.

Let us now consider the size of $G(u,s)$ for fixed $s$. The Gaussian term $e^{- 2 m u^2/s}$ dampens the integrand for increasing $u$. Along with the term in Eq.~\eqref{approxLargeNBar}, $G(u,s)$ is maximized when $u=0$, and the next largest maxima occur when $u=\pm 2 \pi$. Therefore, we can neglect all the components other than $p=0$ in Eq.~\eqref{approxLargeNBar} if the Gaussian term $e^{- 2 m u^2/s}$ is sufficiently small for all $p \neq 0$. This is true when
\begin{equation}
    1 \gg e^{- 8 \pi^2 m/s} 
    \quad \implies \quad
    m > \frac{s}{8 \pi^2}.
\end{equation}
Because the largest relevant value is $s=1/\bar n_0$, this condition is strictest when
\begin{equation}
\label{verySmallM}
    m > m_{min} \equiv \frac{1}{8 \pi^2 \bar n_0}.
\end{equation}
Because we are interested in the thermodynamic limit $\bar n_0 \rightarrow \infty$, we can approximate Eq.~\eqref{approxLargeNBar} with only the $p=0$ component, and our result will be valid for all $m$. Using these results, Eq.~\eqref{fidSum} becomes
\begin{equation}
\label{approxSum}
\begin{aligned}
    \sum_{l,l'\geq0} K_{l,l'} 
        \frac{\bar n_0^{l+l'}}{l!l'!} 
    \approx \frac{e^{2 \bar n_0}}{\bar n_0} \sqrt{2\pi m}e^{2m} \erfc(\sqrt{2m}).
\end{aligned}
\end{equation}

Substituting this result into Eq.~\eqref{conditionalFidelity} and applying the thermodynamic limit $\bar n_0 \rightarrow \infty$, we find that
\begin{equation}
\label{fidelityThermodynamicSM}
    \lim_{\bar n_0 \rightarrow \infty} f(\bar n_0,m)=
        \sqrt{2\pi m}e^{2m} \erfc(\sqrt{2m}).
\end{equation}
We also point out that $F$ is exactly the $n=0$ population of the displaced cavity field [see Eq.~\eqref{displacedCavityState}]:
\begin{equation}
\begin{aligned}
     F \bigl[ \hat \rho_L(\infty),\hat \rho_L(0) \bigr] &\equiv
     \braket{\alpha | \hat \rho_L(\infty) | \alpha } \\
     &= \braket{0 |\hat D(-\alpha) \hat \rho_L(\infty) \hat D (\alpha) |0} \\
     &= \braket{0 | \hat \eta_L |0}.
\end{aligned}
\end{equation}

%%%%%%%%%%%%%%%%%%%%%%%%%%%%%%%%%%%%%%%%%%%%%%%%%%%%%%%%%%%%%%%%%%%%%%%%
\section{Equilibrium quantum entropies and mutual information}
\label{entropyMISMSection}
%%%%%%%%%%%%%%%%%%%%%%%%%%%%%%%%%%%%%%%%%%%%%%%%%%%%%%%%%%%%%%%%%%%%%%%%

In this Appendix we derive expressions for the equilibrium (von Neumann) entropy of various subspaces, as well as the equilibrium quantum mutual information between the cavity field and auxiliary particle $I(\hat \rho_R: \hat \rho_L)$. We first purify the initial particle state $\hat \rho_A(0)$ by using the Schmidt decomposition:
\begin{equation}
    \ket{u}_{AR} = \sum_i \sqrt{\beta_i} \ket{\psi_i}_A \otimes \ket{\phi_i}_R = \sum_i \sqrt{\beta_i} \ket{\psi_i,\phi_i},
\end{equation}
where $\ket{\psi_n}_A$ is an eigenvector of $\hat{\rho}_A(0)$ with eigenvalue $\beta_n$ and $\ket{\phi_n}_R \in R$ for an auxiliary Hilbert space $R$. (Note that we drop the subscript labels whenever it is unambiguous for notational simplicity.) Because we prepare the initial particle state as

\begin{equation}
    \hat \rho_A(0) = x \ket{b}\bra{b} + (1-x)\ket{d}\bra{d},
\end{equation}
then it follows that an appropriate Schmidt decomposition is
\begin{equation}
    \ket{u}_{AR} = \sqrt{x} \ket{b,b} + \sqrt{1-x} \ket{d,d}.
\end{equation}
The corresponding initial (pure) state of the entire $ARL$ (particle + auxiliary particle + cavity) system is then
\begin{widetext}
\begin{equation}
\label{initSystem}
    \hat \rho_{ARL}(0) = 
    \ket{u} \bra{u}_{AR} 
    \otimes 
    \ket{\alpha} \bra{\alpha}= 
    \left\{
        x\ket{b,b}\bra{b,b} 
        + (1-x) \ket{d,d} \bra{d,d} 
        + \sqrt{x(1-x)}
        \left[
            \ket{b,b} \bra{d,d} + \text{h.c.}
        \right]
    \right\} 
    \otimes 
    \ket{\alpha} \bra{\alpha}.
\end{equation}
We now evolve each term in Eq.~\eqref{initSystem} to its corresponding equilibrium state by using Eq.~\eqref{rhoDEq} and Eq.~\eqref{rhoBEq}:
\begin{equation}
    \begin{aligned}
    \hat \rho_{ARL}(\infty) =
    & x \ket{b}\bra{b}_R \otimes 
        \left(
            e^{-\bar n_0} \ket{b} \bra{b}_A \otimes \ket{0}\bra{0}_L
            + \bar n_0 \ket{d} \bra{d}_A \otimes \sum_{l,l' \geq 0} K_{l,l'} \braket{l'|\alpha} \braket{\alpha|l} \op{l'}{l}_L
        \right) \\
    & + (1-x) \ket{d,d}\bra{d,d} \otimes \ket{\alpha}\bra{\alpha}
    + \sqrt{x(1-x)} e^{- \bar n_0/2} \left(
        \ket{b,b}\bra{d,d} \otimes
        \ket{0}\bra{\alpha}_L 
        + \text{h.c.}
    \right).
    \end{aligned}
\end{equation}
\end{widetext}

From this result, the reduced density matrices in the $RL$, $R$, and $L$ subspaces are
\begin{equation}
\label{subspaces}
\begin{aligned}
    \hat \rho_{RL}(\infty) & = x \ket{b} \bra{b} \otimes \hat \rho_c
    + (1-x) \ket{d}\bra{d} \otimes \ket{\alpha}\bra{\alpha}; \\
    \hat \rho_R(\infty) & = x \ket{b} \bra{b} + (1-x) \ket{d} \bra{d}; \\
    \hat \rho_L(\infty) & = x \hat \rho_c + (1-x) \ket{\alpha} \bra{\alpha},
\end{aligned}
\end{equation}
where $\hat \rho_c$ is given in Eq.~\eqref{rhoc}.
Notice that $\hat \rho_{RL}(\infty)$ is separable, and is therefore not an entangled state.
The corresponding von Neumann entropies of the $RL$ and $R$ subspaces are calculated to be
\begin{equation}
\label{VNE}
\begin{aligned}
    S[\hat \rho_{RL}(\infty)] &= S_0 + x S(\hat \rho_c); \\
    S[\hat \rho_R(\infty)] & = S_0,
\end{aligned}
\end{equation}
in which $S_0 \equiv S[\hat \rho_A(0)]$. We do not investigate the behavior of $S(\hat \rho_c)$, but we see from Eq.~\eqref{rhoc} that $\hat \rho_c$ is generally not pure, and hence has nonzero entropy. The von Neumann entropy of the $L$ subspace $S[\hat \rho_L(\infty)]$ is more difficult to calculate because $\hat \rho_c$ and $\ket{\alpha} \bra{\alpha}$ do not always have orthogonal support. 
We can, however, use the identity~\cite{Nielsen}
\begin{equation}
\label{nielsenInequality}
\begin{aligned}
    \sum_i p_i S(\hat \rho_i)
    &\leq
    S \left(
        \sum_i p_i \hat \rho_i
    \right)\\
    &\leq
    \sum_i p_i S(\hat \rho_i)
    - \sum_i p_i \ln p_i
\end{aligned}
\end{equation}
to place an upper and lower bound on the cavity field entropy. 
We emphasize that the upper bound in Eq.~\eqref{nielsenInequality} is only reached when the $\hat \rho_i$'s have orthogonal support.
From Eqs.~\eqref{subspaces} and~\eqref{nielsenInequality},
\begin{equation}
\label{cavityInequality}
    x S(\hat \rho_c) \leq S[\hat \rho_L(\infty)] \leq S_0 + x S(\hat \rho_c).
\end{equation}
We can use this result to place bounds on the equilibrium quantum mutual information between the $R$ and $L$ subspaces:
\begin{equation}
\label{MIRL}
\begin{aligned}
    0 &\leq I[\hat \rho_R(\infty) : \hat \rho_L(\infty)]\\
    &= S[\hat \rho_R(\infty)] + S[\hat \rho_L(\infty)] - S[\hat \rho_{RL}(\infty)]\\
    &\leq S_0.
\end{aligned}
\end{equation}

It is interesting to consider when the equilibrium quantum mutual information $I[\hat \rho_R(\infty) : \hat \rho_L(\infty)]$ is maximized, i.e., when the equilibrium cavity field contains complete information about the initial particle state. From Eqs.~\eqref{subspaces} and~\eqref{nielsenInequality}, this occurs only if the density matrices $\hat \rho_c$ and $\ket{\alpha}\bra{\alpha}$ have orthogonal support. This condition is equivalent to
\begin{equation}
    F(\hat \rho_c,\ket{\alpha}\bra{\alpha})
    = 0,
\end{equation}
where $F$ is the fidelity [see Eq.~\eqref{fidelitySM}]. Using Eqs.~\eqref{finalCavity} and~\eqref{fidEarly}, this can be rewritten as
\begin{equation}
\label{orthogsupportStep}
    f(\bar n_0,m) = 
    e^{-2 \bar n_0}
    \left(
    1 +
    \bar n_0
    \sum_{l,l'\geq 0} K_{l,l'} \frac{\bar n_0^{l+l'}}{l!l'!}
    \right)
    =0.
\end{equation}
Because the term in parentheses is positive, the only way to satisfy Eq.~\eqref{orthogsupportStep} is to consider the thermodynamic limit $\bar n_0 \rightarrow \infty$. In this limit, we can then use Eq.~\eqref{approxSum} to simplify Eq.~\eqref{orthogsupportStep} to
\begin{equation}
       \sqrt{2\pi m}e^{2m} \erfc(\sqrt{2m}) = 0,
\end{equation}
which is true only when $m\rightarrow0$. Therefore, we come to the conclusion that $I(\hat \rho_R: \hat \rho_L) = S_0$, and therefore that the cavity state contains all of the information about the initial particle state, only in the strong-coupling ($m\rightarrow0$), thermodynamic ($\bar n_0 \rightarrow \infty$) limit.

%%%%%%%%%%%%%%%%%%%%%%%%%%%%%%%%%%%%%%%%%%%%%%%%%%%%%%%%%%%%%%%%%
\section{Entanglement with the reservoir}
\label{photonEntropySM}
%%%%%%%%%%%%%%%%%%%%%%%%%%%%%%%%%%%%%%%%%%%%%%%%%%%%%%%%%%%%%%%%%

Here we show that the spontaneous photons in the external reservoir contain at least as much information about the initial particle state as the cavity field.
We also show how this result implies that both the particle and cavity become entangled with the reservoir.

Consider the quantum state describing the particle $A$, cavity field $L$, and external reservoir $P$ (which we model as an infinite bandwidth bosonic bath).
In our Gedankenexperiment, these three subspaces constitute the entire universe, i.e., they form a closed system and therefore undergo unitary evolution.
If we purify the particle state (as done previously) with the auxiliary Hilbert space $R$, then the density matrix of the entire system at any time $t$ is
\begin{equation}
    \hat \rho_{ARLP}(t) = \ket{\Psi_{ARLP}(t)} \bra{\Psi_{ARLP}(t)},
\end{equation}
which is obviously pure.
It follows that the entanglement entropy between the $ARL$ and $P$ subsystems satisfies
\begin{equation}
\label{entanglemententropy}
    S[\hat \rho_{ARL}(t)] = S[\hat \rho_P(t)].
\end{equation}

Suppose we propagate the entire $ARLP$ system to equilibrium.
For $\bar n_0 \gg 1$, the equilibrium particle state $\hat \rho_A(\infty)$ is approximately pure (or completely pure for $\bar n_0 \rightarrow \infty$) because the particle almost certainly ends in the dark state $\ket{d}$:
\begin{equation}
\label{atomApproxPure}
\begin{aligned}
    \hat \rho_A(\infty) & = x e^{- \bar n_0} \ket{b}\bra{b} +
    (1-x e^{- \bar n_0} ) \ket{d} \bra{d} \\
    & \approx \, \ket{d} \bra{d}.
\end{aligned}
\end{equation}
It immediately follows that 
\begin{equation}
\label{separableARL}
    \hat \rho_{ARL}(\infty) \approx \ket{d}\bra{d} \otimes \hat \rho_{RL}(\infty).
\end{equation}
The entanglement entropy is then
\begin{equation}
     S[\hat \rho_{ARL}(\infty)] \approx S[\hat \rho_{RL}(\infty)]
     = S_0 + x S(\hat \rho_c),
\end{equation}
in which we used Eq.~\eqref{VNE}.
Combining this result with Eqs.~\eqref{cavityInequality} and~\eqref{entanglemententropy},
\begin{equation}
\label{spontEntropy}
\begin{aligned}
    S[\hat \rho_L(\infty)] & \leq
    S_0 + x S(\hat \rho_c)\\
    &= S[\hat \rho_{RL}(\infty)] \\
    &\approx 
    S[\hat \rho_{ARL}(\infty)] \\
    &=S[\hat \rho_P(\infty)],
\end{aligned}
\end{equation}
or
\begin{equation}
\label{spontEntropyBigger}
    S[\hat \rho_L(\infty)] \leq S[\hat \rho_P(\infty)].
\end{equation}

Eq.~\eqref{spontEntropyBigger} states that the reservoir $S[\hat \rho_P(\infty)]$ always absorbs at least as much information entropy as the equilibrium cavity state.
Therefore, spontaneous emission in our setting is always a sufficient mechanism for the removal of entropy from the system.

We now show that the reservoir is entangled with both the particle and the cavity by considering the equilibrium quantum mutual information values $I[\hat \rho_P(\infty) : \hat \rho_R(\infty)]$ and $I[\hat \rho_P(\infty) : \hat \rho_L(\infty)]$. 
For simplicity, we restrict ourselves to the large photon number regime $\bar n_0 \gg 1$.
While the reservoir always becomes correlated with the particle and cavity, the entanglement only manifests when Eq.~\eqref{spontEntropyBigger} is a strict inequality, i.e., when $m>0$.
Using Eqs.~\eqref{VNE},~\eqref{atomApproxPure} and the strict inequality version of~\eqref{spontEntropyBigger},
\begin{equation}
\begin{aligned}
    I[\hat \rho_P(\infty) &: \hat \rho_R(\infty)] \\ 
    &=S[\hat \rho_P(\infty)] + S[\hat \rho_R(\infty)] - S[\hat \rho_{PR}(\infty)] \\
    &=S[\hat \rho_P(\infty)] + S[\hat \rho_R(\infty)] - S[\hat \rho_{AL}(\infty)] \\
    &\approx S[\hat \rho_P(\infty)] + S[\hat \rho_R(\infty)] - S[\hat \rho_L(\infty)]\\
    &> S[\hat \rho_R(\infty)] \\
    &\geq \text{min}\left\{ S[\hat \rho_P(\infty)], S[\hat \rho_R(\infty)] \right\},
\end{aligned}
\end{equation}
which shows that the reservoir is entangled with the initial particle state. Similarly, using Eqs.~\eqref{VNE},~\eqref{atomApproxPure}, and the strict inequality version of Eq.~\eqref{spontEntropy},
\begin{equation}
\begin{aligned}
    I[\hat \rho_P(\infty) &: \hat \rho_L(\infty)] \\
    &=S[\hat \rho_P(\infty)] + S[\hat \rho_L(\infty)] - S[\hat \rho_{PL}(\infty)] \\
    &=S[\hat \rho_P(\infty)] + S[\hat \rho_L(\infty)] - S[\hat \rho_{AR}(\infty)] \\
    &\approx S[\hat \rho_P(\infty)] + S[\hat \rho_L(\infty)] - S[\hat \rho_R(\infty)]\\
    &> S[\hat \rho_L(\infty)] \\
    &\geq \text{min}\left\{ S[\hat \rho_P(\infty)], S[\hat \rho_L(\infty)] \right\},
\end{aligned}
\end{equation}
which shows that the reservoir is entangled with the final cavity state.

\bibliography{references.bib}

\end{document}